\documentclass{article}
\usepackage{spconf,amsmath,graphicx,hyperref}
\usepackage{adjustbox}
\usepackage{multirow}
\usepackage{tabularx}
\usepackage{threeparttable,booktabs,amssymb}
\usepackage[numbers,sort&compress]{natbib}

\title{MQAScore: Fine-Grained Text-Music Alignment Evaluation\\
via Large Audio Language Model}
\name{Wei-Jaw Lee$^{1, 2, 3}$, Chun-Ping Wang$^{1}$, Xuanjun Chen$^{1, 2}$, Yi-Hsuan Yang$^{1, 2, 3}$}
\address{$^{1}$GICE, National Taiwan University;\ $^{2}$AI-CoRE, National Taiwan University;\ $^{3}$Taiwan AILabs}
\begin{document}
%
\maketitle
\begin{abstract} 
Text-to-music alignment is commonly evaluated by the global similarity between text and audio embeddings, which collapses all described attributes into a single score and obscures which ones are missing. We instead introduce MQAScore, which verifies each atomic musical concept along five clip-level dimensions (genre, instrument, mood/theme, vocal, and usage) by the probability that a large audio language model (LALM) answers affirmatively to a binary question. To assess evaluator reliability, we construct verification labels from presence/absence annotations and LLM-guided concept substitutions, and propose the Sensitivity-Invariance Ratio, which measures how selectively an evaluator responds to meaning-changing substitutions. Experiments show that concept-level scoring improves both question-answering and similarity-based evaluators, and MQAScore can be a more reliable and interpretable fine-grained complement to global similarity for evaluating text-to-music alignment. 
\end{abstract}
\section{Introduction}
\label{sec:intro}

\begin{figure}
  \centering
  \includegraphics[width=0.8\columnwidth]
  {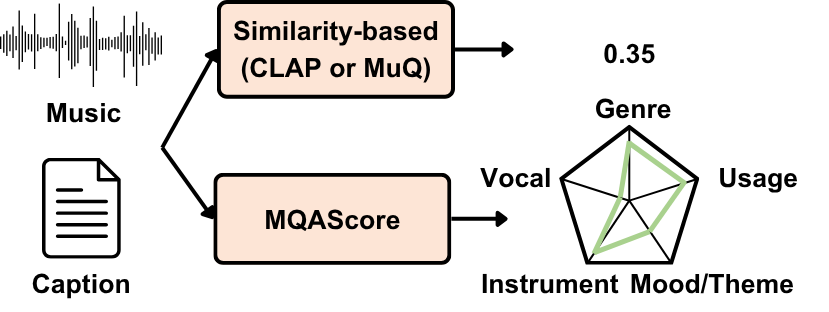}
  \caption{Global similarity (e.g. CLAP, MuQ-MuLan) yields a single score; MQAScore verifies each musical dimension separately.}
  \label{fig:differ}
\end{figure}

The field of music generation has evolved rapidly, with state-of-the-art text-to-music (TTM) systems now capable of generating high-fidelity and vocalized music.
Alongside these advancements, text remains the primary conditioning modality, with prompts becoming more intricate to improve controllability. To evaluate text-audio alignment, existing studies predominantly rely on similarity-based metrics~\cite{wu2023laionclap, zhu2025muq} computing a global similarity score between text and audio embeddings.  However, as shown in Fig.~\ref{fig:differ}, a global similarity score summarizes all attributes in the prompt and audio into a single value, obscuring fine-grained semantic information and making the evaluation difficult to interpret. As text-to-music generation moves toward detailed, multi-attribute prompts, alignment evaluation should move beyond global similarity and focus on evaluating individual semantic concepts.

Moving toward such fine-grained concept-level evaluation raises two challenges. The first is to define the evaluation space, as musical captions span attributes ranging from descriptive (e.g., genre, instrument) to musicological (e.g., key, chord, tempo) ones. 
To address this challenge, we ground our taxonomy in two tagging schemes. CAL500~\cite{turnbull2008semantic}, which spans six dimensions: emotion, genre, instrument, vocal, usage, and acoustic qualities; MTG-Jamendo~\cite{bogdanov2019mtg} instead groups affective and evocative tags into a single \emph{mood/theme} category. 
We retain the first five CAL500 dimensions and follow the latter in naming the first one mood/theme, since listener-facing tags rarely separate how a clip feels from what it evokes; we reserve usage for the listening situation a clip is intended for. Their concepts are categorical, so the presence or absence of each concept in a clip is well defined, although detection difficulty varies across dimensions~\cite{ma2025cmibench}.
We exclude acoustic qualities, which reflect audio fidelity rather than semantic alignment. Moreover, we also exclude musicological concepts such as those defined by the Pandora's Music Genome Project~\cite{pandora_mgp}
(e.g., harmony, rhythm), which remain difficult to detect by LALMs~\cite{li2026mrmad, weck2024muchomusic}.

The second challenge is cost: overall human ratings of text-to-music systems already require thousands of annotators and comparisons~\cite{solak2024benchmarking}, and concept-level annotation multiplies this cost by the number of attributes per prompt. LALMs offer a scalable alternative given their rapidly improving music understanding capabilities~\cite{xu2025qwen3omni, ghosh2026musicflamingo, ghosh2026afnext}, yet remain unreliable at open-ended musical reasoning~\cite{weck2024muchomusic, weck2026hummusqa}. Following recent text-to-audio alignment metrics~\cite{kuan2026aqascore, li2026anyaudio, hsieh2026attm}, we restrict the LALM to deciding whether a single concept is supported by the clip. 
Specifically, the proposed ``MQAScore'' adopts the scoring mechanism of AQAScore~\cite{kuan2026aqascore}, the probability of an affirmative answer, but applies it to individual concepts rather than to the full prompt.

Since captions reveal which concepts are present but not which are absent, we obtain negatives from presence/absence annotations for genre and instrument, and for the other three dimensions by replacing each concept with its most distant alternative under an LLM's text prior, using the closest alternative as a meaning-preserving control. To quantify the contrast, we introduce the Sensitivity-Invariance Ratio ($R_{S/I}$), which measures whether score changes follow semantic contradictions rather than minor textual edits.
We then assess MQAScore by how much more it responds to contradicting substitutes than to meaning-preserving ones, and by its agreement with experts in pairwise comparisons.

In summary, our contributions are threefold: 1) We introduce a taxonomy of five clip-level dimensions grounded in CAL500, together with MQAScore, which verifies each atomic musical concept separately with an LALM.
2) We construct verification labels from presence/absence annotations and LLM-guided concept substitutions, and propose the Sensitivity-Invariance Ratio ($R_{S/I}$) to measure how selectively an evaluator responds to meaning-changing substitutions.
3) We show that concept-level decomposition improves both similarity-based and question-answering-based scores, and that LALM-based scoring further outperforms similarity-based scoring in both discrimination and human agreement.

Due to space limits, the paper reports verification results that assess the reliability of the evaluators themselves. To demonstrate how MQAScore applies in practice, we additionally provide non-cherry-picked per-dimension evaluation results of state-of-the-art TTM systems, together with the audio clips, the extracted concepts, and the per-concept scores behind each dimension, on the project page: \url{https://lonian6.github.io/MQAScore.github.io/}.
These results illustrate the diagnostic value of per-dimension scoring: some of the systems we evaluate generate vocals and others do not, a difference that global similarity barely reflects but that MQAScore isolates in the vocal dimension.





\section{Related Work}
\label{sec:relatedwork}


Despite the rapid advancement of music generation models, automatic evaluation of text-music alignment remains challenging. Conventional text-audio and text-music alignment methods largely rely on global similarity in joint cross-modal embedding spaces, such as CLAP-based scores, and MuQ-MuLan~\cite{elizalde2023clap, wu2023laionclap, zhu2025muq}. Studies have highlighted the limitations of such global representations in capturing detailed musical attributes~\cite{anand2026figma}, while finer-grained evaluation methods have emerged to assess individual acoustic events, musical attributes, or compositional constraints~\cite{suzuki2026elsa, ma2026cmirewardbench}. Evaluation research has also emphasized stronger agreement with human judgments through subjective alignment ratings, expert assessments, mean opinion scores, or preference annotations~\cite{solak2024benchmarking, liu2025musiceval, takano2025humanclap, okamoto2026xacle, wang2026deramos, yang2026fusemos, ma2026cmirewardbench}. 
Meanwhile, music question-answering datasets and benchmarks have been developed to probe LALM capabilities in music understanding, instruction following, and reasoning~\cite{liu2024mullama, deng2024musilingo, weck2024muchomusic, ma2025cmibench}, while recent general and music-specialized LALMs, including Audio Flamingo Next, Music Flamingo, and Qwen3-Omni, demonstrate increasingly strong capabilities in these areas~\cite{ghosh2026afnext, ghosh2026musicflamingo, xu2025qwen3omni}. Building on these capabilities, emerging approaches have begun repurposing LALMs themselves as semantic evaluators, reformulating audio–text alignment as question answering, rubric-based verification, or concept-level judging~\cite{kuan2026aqascore, li2026anyaudio, wang2026audioscape, hsieh2026attm}.

\section{Method}
\label{sec:method}


\subsection{Taxonomy Construction}
\label{sec:taxonomy}


Following CAL500~\cite{turnbull2008semantic}, we adopt five dimensions as 
\begin{equation*}
    \mathcal{D} = \{\mathrm{Genre}, \mathrm{Instrument}, \mathrm{Mood/Theme}, \mathrm{Vocal}, \mathrm{Usage}\}.
\end{equation*}
For each caption $c$ in MusicCaps~\cite{agostinelli2023musiclm}, an LLM extracts clip-level concepts $\mathcal{E}(c) = \{(d_i, t_i)\}_{i=1}^{K}$, assigning each atomic concept $t_i$ to a dimension $d_i \in \mathcal{D}$ and discarding subjective modifiers (e.g., ``groovy bass'' yields ``bass''). The vocabulary $\mathcal{V}_d$ collects the concepts extracted for dimension $d$ (e.g., bass $\in \mathcal{V}_{\mathrm{Instrument}}$), after merging synonyms and manual review.


\noindent
\textbf{Verification Label Construction.}
\label{sec:labels}
Verifying a concept requires both positives, i.e., concepts supported by the clip, and negatives, i.e., concepts not supported by it. Given a clip $a$ and a concept $t \in \mathcal{V}_d$ supported by $a$, we construct negatives $t^- \in \mathcal{V}_d$ in a dimension-dependent manner, and in all dimensions, a reliable evaluator should score $t$ above $t^-$. For genre and instrument, we pair $t$ with each concept annotated as absent from $a$~\cite{tzanetakis2002gtzan, humphrey2018openmic}, forming triplets $(a, t, t^-)$.
For mood/theme, vocal, and usage, such annotations are unavailable, and concepts are not mutually exclusive (e.g., a clip can be both calm and melancholic), so an arbitrary substitute may still be supported. We therefore replace $t$ with its semantically most distant concept in $\mathcal{V}_d$, as judged by an LLM, to obtain $t^-$. Since this substitution alters the text itself, a lower score for $t^-$ may reflect the edit rather than the audio; we thus also take the closest concept as a meaning-preserving substitute $t^+ \in \mathcal{V}_d$, forming quadruplets $(a, t, t^+, t^-)$. For these dimensions, a reliable evaluator should additionally respond little to $t^+$; since $t^+$ and $t^-$ undergo the same substitution, a stronger response to $t^-$ than to $t^+$ indicates that the score change stems from the audio rather than the edit. See Sec.~\ref{sec:metrics} for how we utilize the triplets and quadruplets for verification.

\subsection{Concept Verification Scoring}
\label{sec:scoring}

Following AQAScore~\cite{kuan2026aqascore}, we cast evaluation as audio-grounded binary question answering, but query each concept separately rather than the full prompt. For a concept $t \in \mathcal{V}_d$, a dimension-specific template $\phi_d$ produces a question $q = \phi_d(t)$ (e.g., {\texttt{Does this music belong to the genre \{t\}? Answer Yes or No.}), and an LALM $p_\theta$ scores a clip $a$ by the probability of ``Yes'' normalized against ``No'':
\begin{equation*}
    S(a, t) = \frac{p_\theta(\text{``Yes''} \mid a, q)}{p_\theta(\text{``Yes''} \mid a, q) + p_\theta(\text{``No''} \mid a, q)},
\end{equation*}
which we refer to as MQAScore. Given the concepts $\mathcal{E}(c) = \{(d_i, t_i)\}_{i=1}^{K}$ extracted from a caption $c$, the alignment $S_d(a, c)$ along dimension $d \in \mathcal{D}$ is the mean of $S(a, t_i)$ over $\mathcal{I}_d = \{i \mid d_i = d\}$, yielding a per-dimension score vector rather than a single global score; dimensions with $\mathcal{I}_d = \emptyset$ are reported as unavailable rather than assigned zero.

\section{Experiments}
\label{sec:exp}

\subsection{Verification Data}

\textbf{Genre and Instrument.} GTZAN~\cite{tzanetakis2002gtzan} contains 1,000 30-second clips evenly distributed over 10 genres. As GTZAN is single-labeled, we take each clip's genre as $t$ and treat the remaining nine genres as negatives $t^-$, yielding nine triplets $(a, t, t^-)$ per clip.
OpenMIC-2018~\cite{humphrey2018openmic} provides 20,000 10-second clips with partial annotations over 20 instruments, including explicit absence. Following the official release, we label an annotated instrument present if its relevance exceeds 0.5 and absent otherwise. Pairing each present instrument with every absent one in the same clip, excluding voice, yields 15,241 triplets over 7,037 clips.

\noindent
\textbf{Mood/Theme, Vocal, and Usage.} We apply the concept extraction in Sec.~\ref{sec:taxonomy} on MusicCaps~\cite{agostinelli2023musiclm} and construct $t^+$ and $t^-$ for each extracted concept yielding 3967, 737, and 112 quadruplets $(a, t, t^+, t^-)$ for mood/theme, vocal, and usage, respectively. The LLM we used for concept extraction and constructing $t^+$ and $t^-$ is \texttt{Qwen3-4B-Instruct-2507}.

\noindent
\textbf{Expert Pairwise Annotations.} To test whether MQAScore ranks pairs as experts do, we sample clips from MusicCaps and have  11 annotators with musical training label pairs following written guidelines. For genre and instrument, given a concept $t$ and two clips $(a_1, a_2)$, annotators label which clip exhibits $t$. For mood/theme, vocal, and usage, given a clip $a$ and a pair $(t, t^-)$, annotators label which concept applies to $a$. The label set also includes ``both'' and ``neither''; we retain only pairs with an unambiguous label, yielding 147 and 119 pairs for the two types, respectively, with an inter-annotator agreement of Krippendorff's $\alpha$ = 0.71.


\subsection{Evaluators Assessed}
\label{sec:evaluators}
We evaluate two scoring paradigms at two input granularities. At the \textit{caption-level}, an evaluator scores the full caption $c$, and a substitute is evaluated by replacing $t$ with $t^{+}$ or $t^{-}$ within $c$. At the \textit{concept-level}, each extracted concept is scored in isolation as $S(a, t)$. Caption-level scoring applies only to mood/theme, vocal, and usage, as GTZAN and OpenMIC-2018 provide tag annotations rather than captions.

\noindent
\textbf{Similarity-based.} We use two contrastive text-audio models, LAION-CLAP~\cite{wu2023laionclap} (\texttt{music\_audioset\_epoch\_15 \_esc\_90.14.pt}) and MuQ-MuLan~\cite{zhu2025muq} (\texttt{OpenMuQ/MuQ- MuLan-large}), and take the cosine similarity between the text and audio embeddings as the score. At the concept level, each concept is embedded with the template ``\texttt{This is a music piece about \{t\}.}''

\noindent
\textbf{QA-based.} We use three LALMs: Qwen3-Omni~\cite{xu2025qwen3omni} (\texttt{Qwen3 -Omni-30B-A3B-Instruct}), Audio Flamingo Next~\cite{ghosh2026afnext}, and Music Flamingo~\cite{ghosh2026musicflamingo}. At the caption level, we apply AQAScore~\cite{kuan2026aqascore}, which asks whether the clip matches the full caption; at the concept level, we apply MQAScore. Scores are computed from the next-token probabilities of ``Yes'' and ``No'' in a single forward pass, without sampling.
\begin{table*}[t]
\centering
\small
\setlength{\tabcolsep}{5pt}
\caption{Concept verification results: AUC with WR for genre and instrument, and with $R_{S/I}$ for mood/theme, vocal, and usage. Caption-level scores are unavailable for tag-only genre and instrument data.}
\label{tab:main}
\begin{tabular}{lll cc cc c cc cc cc}
\toprule
\multirow{2}{*}{Input} & \multirow{2}{*}{Method} & \multirow{2}{*}{Model}
 & \multicolumn{2}{c}{Genre} & \multicolumn{2}{c}{Instrument}
 & & \multicolumn{2}{c}{Mood/Theme} & \multicolumn{2}{c}{Vocal}
 & \multicolumn{2}{c}{Usage} \\
\cmidrule(lr){4-5}\cmidrule(lr){6-7}
\cmidrule(lr){9-10}\cmidrule(lr){11-12}\cmidrule(lr){13-14}
 & & & AUC & WR & AUC & WR
 & & AUC & $R_{S/I}$ & AUC & $R_{S/I}$ & AUC & $R_{S/I}$ \\
\midrule
\multirow{5}{*}{Caption}
 & \multirow{3}{*}{AQAScore}
   & Qwen3-Omni     & -- & -- & -- & -- & & 0.776 & 31.3 & 0.870 & 31.0 & 0.718 & 36.9 \\
 & & AF-Next        & -- & -- & -- & -- & & 0.739 & 20.6 & 0.727 &  9.4 & 0.685 &  6.9 \\
 & & Music Flamingo & -- & -- & -- & -- & & 0.684 & 10.0 & 0.711 &  6.8 & 0.610 &  5.1 \\
\cmidrule(lr){2-14}
 & \multirow{2}{*}{Similarity}
   & CLAP           & -- & -- & -- & -- & & 0.535 & 1.94 & 0.629 & 2.63 & 0.559 & 1.47 \\
 & & MuQ-MuLan            & -- & -- & -- & -- & & 0.532 & 2.34 & 0.500 & 0.38 & 0.528 & 1.55 \\
\midrule
\multirow{5}{*}{Concept}
 & \multirow{3}{*}{MQAScore}
   & Qwen3-Omni     & {0.983} & {0.978} & {0.915} & {0.887} & & 0.943 & 7.38 & 0.966 & 21.90 & 0.842 & 9.09 \\
 & & AF-Next        & 0.980 & 0.972 & 0.893 & 0.849 & & 0.900 & 5.12 & 0.881 &  7.03 & 0.744 & 4.43 \\
 & & Music Flamingo & 0.982 & 0.977 & 0.855 & 0.797 & & 0.885 & 3.22 & 0.886 &  3.28 & 0.713 & 3.03 \\
\cmidrule(lr){2-14}
 & \multirow{2}{*}{Similarity}
   & CLAP           & 0.912 & 0.930 & 0.867 & 0.849 & & 0.669 & 1.91 & 0.759 & 2.06 & 0.711 & 1.39 \\
 & & MuQ-MuLan            & 0.959 & 0.946 & 0.860 & 0.812 & & 0.663 & 1.73 & 0.511 & 0.48 & 0.654 & 1.53 \\
\bottomrule
\end{tabular}
\end{table*}

\subsection{Reliability Metrics}
\label{sec:metrics}

\textbf{AUC.} For each dimension $d$, we pool the scores of all positives $S(a, t)$ and negatives $S(a, t^-)$ across clips and report the unpaired ROC-AUC (Mann-Whitney $U$), with ties counted as 0.5. As positives and negatives are not paired within a clip, it also penalizes evaluators whose scores drift across clips.

\noindent
\textbf{Win Rate (WR).} For genre and instrument, we further measure within-clip discrimination. For each clip, we compute the fraction of (present, absent) concept pairs in which the present concept is scored higher, and average this fraction over clips with at least one present and one absent annotation.

\noindent
\textbf{Sensitivity-Invariance Ratio ($R_{S/I}$).} 
For mood/theme, vocal, and usage, we measure how much scores change under substitution by the normalized mean absolute difference ($z$-MAD) over quadruplets $(a, t, t^+, t^-)$:
\begin{equation*}
    z\text{-MAD}^{k} = \mathbb{E}_{(a, t, t^{k})}\Big[\big|\tilde{S}(a, t) - \tilde{S}(a, t^{k})\big|\Big], \quad k \in \{-, +\},
\end{equation*}
where $\tilde{S}(a, t) = (S(a, t) - \mu_S)/\sigma_S$ is the z-normalized score, with $\mu_S$ and $\sigma_S$ the mean and standard deviation of each evaluator's scores on the original concepts. We then define $R_{S/I} = z\text{-MAD}^{-} / z\text{-MAD}^{+}$, which quantifies how much further an evaluator's score moves when the concept is contradicted than when it is replaced by a close alternative; $R_{S/I} \gg 1$ indicates that its judgments follow the audio rather than the wording, whereas $R_{S/I} \approx 1$ means it cannot tell a contradiction from a near-synonym.

\noindent
\textbf{Agreement with Expert Labels.} 
Expert labels serve as ground truth. Each evaluator predicts the higher-scoring option, $\hat{y}=\arg\max_{a'\in\{a_1,a_2\}}S(a',t)$ for genre and instrument and $\hat{y}=\arg\max_{t'\in\{t,t^-\}}S(a,t')$ for the other three dimensions, and agreement is the fraction of pairs it matches.

\section{Results}
\label{sec:results}

\noindent
\textbf{Concept Verification Results.}
Table~\ref{tab:main} shows that concept-level scoring improves both paradigms over caption-level scoring, as isolating each concept prevents its contribution from being diluted by the other concepts in a caption. At the concept level, LALMs further outperform similarity-based evaluators on mood/theme, vocal, and usage in both AUC and $R_{S/I}$. Similarity-based evaluators respond almost as strongly to close alternatives as to contradictions, with MuQ-MuLan even reacting more to close alternatives on vocal, suggesting that their scores track the wording rather than the audio. On genre and instrument, similarity-based evaluators remain competitive, likely because these closed vocabularies are well covered by contrastive pretraining. Caption-level AQAScore shows a high $R_{S/I}$ despite its lower AUC, as replacing a single concept in a multi-concept caption barely moves the score under a close alternative, inflating the ratio. Usage is the most challenging dimension for all evaluators, as situational context is inferred from, rather than directly heard in, the audio. Finally, evaluators with similar AUC can differ markedly in $R_{S/I}$, indicating that discrimination accuracy alone does not capture how selectively an evaluator responds to meaning-changing substitutions.

\begin{table}[t]
\centering 
\small     
\renewcommand{\arraystretch}{0.95} 
\setlength{\tabcolsep}{4.3pt}

\caption{Agreement (\%) with expert labels on MusicCaps pairs. H: human-human agreement; Q3-O: Qwen3-Omni; AF-N: Audio Flamingo Next; MF: Music Flamingo; MuQ: MuQ-MuLan. All evaluators use concept-level scoring.}
\label{tab:human}

\begin{tabular}{lcccccc}
\toprule
 & H & Q3-O & AF-N & MF & CLAP & MuQ \\
\midrule
Genre           & 92.5 & 85.0 & 86.7 & 75.0 & 71.7 & 71.7 \\
Instrument      & 82.3 & 79.7 & 72.9 & 89.8 & 74.6 & 62.7 \\
\midrule
Mood/Theme      & 84.6 & 86.4 & 88.6 & 88.6 & 54.5 & 43.2 \\
Vocal           & 95.2 & 92.5 & 92.5 & 85.0 & 80.0 & 60.0 \\
Usage         & 82.5 & 85.4 & 78.0 & 68.3 & 63.4 & 61.0 \\
\bottomrule
\end{tabular}
\end{table}

\noindent
\textbf{Agreement with Expert Labels.}
Table~\ref{tab:human} compares evaluators with expert labels, using human-human agreement as a reference. LALM-based evaluators reach agreement comparable to that among humans across dimensions, whereas similarity-based evaluators fall far behind, most notably on mood/theme, where MuQ-MuLan drops below chance. As in Table~\ref{tab:main}, the gap narrows on genre and instrument, whereas abstract concepts such as mood and usage benefit from the LLM backbone of LALMs, which interprets concept semantics beyond what a single embedding match captures, and from explicitly querying the clip for each concept. Since all evaluators here score concepts individually, this advantage stems from question answering rather than granularity alone. Although rankings among LALMs vary, the gap between LALMs and similarity-based evaluators is consistent.

\section{Conclusion}
\label{sec:conclusion}
We presented MQAScore, which evaluates text-to-music alignment by verifying atomic musical concepts across five dimensions with LALMs. Concept-level scoring improves both QA- and similarity-based evaluators over caption-level scoring, while LALMs outperform similarity-based evaluators and approach human agreement on mood/theme, vocal, and usage. Our Sensitivity-Invariance Ratio further reveals differences in selectivity to meaning-changing substitutions, with usage remaining the most challenging dimension. Limitations include LLM-based concept extraction and negative selection, the single-label assumption of GTZAN, the focus on clip-level descriptive concepts, and the small expert set. Future work will extend the taxonomy to relational concept graphs and musicological attributes, and apply concept verification to generated music and text-to-music reward modeling.

\section{Compliance with Ethical Standards}
This research uses open-access datasets (MusicCaps, GTZAN, and
OpenMIC-2018) in accordance with their licenses. The expert pairwise labels were collected as dataset annotations describing audio clips, not as a study of the annotators themselves. Annotators participated voluntarily, and no personally identifiable or sensitive information about them was collected or analyzed.

\section{Acknowledgments}
The work is supported by grants the National Science and Technology Council of Taiwan (NSTC 114-2628-E-002-013-MY3), and the Ministry of Education (MOE) of Taiwan (for Taiwan Centers of Excellence in Artificial Intelligence). 
Gemini, ChatGPT, and Claude were used solely for language editing
of the manuscript; all edits were reviewed by the authors, who takefull responsibility for the content. The LLM and LALM used as part of the proposed method is described in Sec.~\ref{sec:method}.

\def\thebibliography#1{\section{References}\footnotesize\list
 {[\arabic{enumi}]}{\settowidth\labelwidth{[#1]}\leftmargin\labelwidth
 \advance\leftmargin\labelsep
 \itemsep=0pt plus 0.3ex\relax
 \parsep=0pt\relax
 \usecounter{enumi}}
 \def\newblock{\hskip .11em plus .33em minus .07em}
 \sloppy\clubpenalty4000\widowpenalty4000
 \sfcode`\.=1000\relax}
\bibliographystyle{IEEEbib}
\bibliography{refs}

@inproceedings{elizalde2023clap,
  author={Elizalde and others},
  booktitle={ICASSP 2023 - 2023 IEEE International Conference on Acoustics, Speech and Signal Processing (ICASSP)}, 
  title={CLAP Learning Audio Concepts from Natural Language Supervision}, 
  year={2023},
  volume={},
  number={},
  pages={1-5},
  doi={10.1109/ICASSP49357.2023.10095889}
}

@inproceedings{wu2023laionclap,
  author={Wu, Yusong and others},
  booktitle={ICASSP 2023 - 2023 IEEE International Conference on Acoustics, Speech and Signal Processing (ICASSP)}, 
  title={Large-Scale Contrastive Language-Audio Pretraining with Feature Fusion and Keyword-to-Caption Augmentation}, 
  year={2023},
  volume={},
  number={},
  pages={1-5},
  doi={10.1109/ICASSP49357.2023.10095969}
}

@article{agostinelli2023musiclm,
      title={MusicLM: Generating Music From Text}, 
      author={Andrea Agostinelli and others},
      year={2023},
      eprint={2301.11325},
      archivePrefix={arXiv},
      primaryClass={cs.SD},
      url={https://arxiv.org/abs/2301.11325}, 
      journal={arXiv preprint arXiv:2301.11325},
}

@article{zhu2025muq,
  author={Zhu, Haina and others},
  journal={IEEE Transactions on Audio, Speech and Language Processing}, 
  title={MuQ: Self-Supervised Music Representation Learning With Mel Residual Vector Quantization}, 
  year={2025},
  volume={33},
  number={},
  pages={3653-3664},
  doi={10.1109/TASLPRO.2025.3602320}
}

@inproceedings{anand2026figma,
  title={FIGMA: Towards FIne-Grained Music retrievAl},
  author={Anand, Nishit and others},
  booktitle={Proceedings of the 64th Annual Meeting of the Association for Computational Linguistics (Volume 1: Long Papers)},
  pages={47559--47572},
  year={2026}
}

@article{suzuki2026elsa,
      title={ELSA: Acoustic Event-Level Semantic Alignment for Fine-Grained Reference-Free Text-to-Audio Evaluation}, 
      author={Shuntaro Suzuki and others},
      year={2026},
      eprint={2606.17404},
      archivePrefix={arXiv},
      primaryClass={eess.AS},
      url={https://arxiv.org/abs/2606.17404},
      journal={arXiv preprint arXiv:2606.17404},
}

@article{ma2026cmirewardbench,
      title={CMI-RewardBench: Evaluating Music Reward Models with Compositional Multimodal Instruction}, 
      author={Yinghao Ma and others},
      year={2026},
      eprint={2603.00610},
      archivePrefix={arXiv},
      primaryClass={cs.SD},
      url={https://arxiv.org/abs/2603.00610}, 
      journal={arXiv preprint arXiv:2603.00610},
}

@inproceedings{solak2024benchmarking,
  author={Grötschla, Florian and others},
  booktitle={ICASSP 2025 - 2025 IEEE International Conference on Acoustics, Speech and Signal Processing (ICASSP)}, 
  title={Benchmarking Music Generation Models and Metrics via Human Preference Studies}, 
  year={2025},
  volume={},
  number={},
  pages={1-5},
  doi={10.1109/ICASSP49660.2025.10887745}
}

@inproceedings{liu2025musiceval,
  author={Liu, Cheng and others},
  booktitle={ICASSP 2025 - 2025 IEEE International Conference on Acoustics, Speech and Signal Processing (ICASSP)}, 
  title={MusicEval: A Generative Music Dataset with Expert Ratings for Automatic Text-to-Music Evaluation}, 
  year={2025},
  volume={},
  number={},
  pages={1-5},
  doi={10.1109/ICASSP49660.2025.10890307}
}

@inproceedings{takano2025humanclap,
  author={Takano, Taisei and others},
  booktitle={2025 Asia Pacific Signal and Information Processing Association Annual Summit and Conference (APSIPA ASC)}, 
  title={Human-CLAP: Human-perception-based Contrastive Language-audio Pretraining}, 
  year={2025},
  volume={},
  number={},
  pages={131-136},
  doi={10.1109/APSIPAASC65261.2025.11249121}
}

@inproceedings{okamoto2026xacle,
  author={Okamoto, Yuki and others},
  booktitle={ICASSP 2026 - 2026 IEEE International Conference on Acoustics, Speech and Signal Processing (ICASSP)}, 
  title={XACLE Challenge 2026: The First X-to-Audio Alignment Challenge}, 
  year={2026},
  volume={},
  number={},
  pages={21877-21879},
  doi={10.1109/ICASSP55912.2026.11464875}
}

@article{wang2026deramos,
  author={Wang, Chien-Chun and others},
  journal={IEEE Signal Processing Letters}, 
  title={DeRA-MOS: Optimizing Text-to-Music Evaluation via Decoupled Listwise Ranking and Modality Alignment}, 
  year={2026},
  volume={33},
  number={},
  pages={2740-2744},
  doi={10.1109/LSP.2026.3702515}
}

@inproceedings{yang2026fusemos,
  author={Yang, Jing and others},
  booktitle={ICASSP 2026 - 2026 IEEE International Conference on Acoustics, Speech and Signal Processing (ICASSP)}, 
  title={FUSEMOS: Perceptual Evaluation of Text-to-Music Generation with Dual-Encoder Fusion and Ranking-Aware Composite Loss}, 
  year={2026},
  volume={},
  number={},
  pages={15157-15161},
  doi={10.1109/ICASSP55912.2026.11461672}
}

@inproceedings{liu2024mullama,
  author={Liu, Shansong and others},
  booktitle={ICASSP 2024 - 2024 IEEE International Conference on Acoustics, Speech and Signal Processing (ICASSP)}, 
  title={Music Understanding LLaMA: Advancing Text-to-Music Generation with Question Answering and Captioning}, 
  year={2024},
  volume={},
  number={},
  pages={286-290},
  doi={10.1109/ICASSP48485.2024.10447027}
}

@inproceedings{deng2024musilingo,
  title={Musilingo: Bridging music and text with pre-trained language models for music captioning and query response},
  author={Deng, Zihao and others},
  booktitle={Findings of the Association for Computational Linguistics: NAACL 2024},
  pages={3643--3655},
  year={2024}
}

@inproceedings{weck2024muchomusic,
        title={MuChoMusic: Evaluating Music Understanding in Multimodal Audio-Language Models},
        author={Weck, Benno and others},
        booktitle = {Proceedings of the 25th International Society for Music Information Retrieval Conference (ISMIR)},
        year={2024}
}

@inproceedings{ma2025cmibench,
  title     = {{CMI-Bench}: A Comprehensive Benchmark for Evaluating Music Instruction Following},
  author    = {Ma, Yinghao and others},
  booktitle = {Proceedings of the 26th International Society for Music Information Retrieval Conference (ISMIR)},
  year      = {2025}
}

@article{ghosh2026afnext,
  title={Audio flamingo next: Next-generation open audio-language models for speech, sound, and music},
  author={Ghosh, Sreyan and others},
  journal={arXiv preprint arXiv:2604.10905},
  year={2026}
}

@inproceedings{ghosh2026musicflamingo,
 author = {Ghosh, Sreyan and others},
 booktitle = {International Conference on Learning Representations},
 editor = {C. Vondrick and B. Hariharan and C. Raffel and L. Pinto and D. Yang and A. Faust},
 pages = {78958--78996},
 title = {Music Flamingo: Scaling Music Understanding in Audio Language Models},
 url = {https://proceedings.iclr.cc/paper_files/paper/2026/file/7f64034009f4a5fa417a57e1a987c5cd-Paper-Conference.pdf},
 volume = {2026},
 year = {2026}
}

@article{xu2025qwen3omni,
  title={Qwen3-omni technical report},
  author={Xu, Jin and others},
  journal={arXiv preprint arXiv:2509.17765},
  year={2025}
}

@article{kuan2026aqascore,
  title   = {AQAScore: Evaluating Semantic Alignment in Text-to-Audio Generation via Audio Question Answering},
  author  = {Kuan, Chun-Yi and Chang, Kai-Wei and Lee, Hung-yi},
  journal = {arXiv preprint arXiv:2601.14728},
  year    = {2026},
  url     = {https://arxiv.org/abs/2601.14728},
}

@article{li2026anyaudio,
      title={AnyAudio-Judge: A Dynamic Rubric-Based Benchmark and Evaluator for Audio Instruction Following}, 
      author={Haitao Li and others},
      year={2026},
      eprint={2606.03116},
      archivePrefix={arXiv},
      primaryClass={eess.AS},
      url={https://arxiv.org/abs/2606.03116}, 
      journal={arXiv preprint arXiv:2606.03116},
}

@article{wang2026audioscape,
      title={AudioScape-TTA: A Structured Soundscape Benchmark for Fine-Grained Text-to-Audio Evaluation}, 
      author={Jinting Wang and others},
      year={2026},
      eprint={2608.04479},
      archivePrefix={arXiv},
      primaryClass={cs.SD},
      url={https://arxiv.org/abs/2608.04479},
      journal={arXiv preprint arXiv:2608.04479},
}

@inproceedings{hsieh2026attm,
  author    = {Fang-Chih Hsieh and others},
  title     = {Academic Text-to-Music Grand Challenge:
               Datasets, Baselines, and Evaluation Methods},
  booktitle = {2026 IEEE International Conference on Multimedia and Expo (ICME)},
  year      = {2026},
  note      = {Grand Challenge Paper}
}

@conference {bogdanov2019mtg,
    author = "Bogdanov, Dmitry and others",
    title = "The MTG-Jamendo Dataset for Automatic Music Tagging",
    booktitle = "Machine Learning for Music Discovery Workshop, International Conference on Machine Learning (ICML 2019)",
    year = "2019",
    address = "Long Beach, CA, United States",
    url = "http://hdl.handle.net/10230/42015"
}

@inproceedings{weck2026hummusqa,
  title={HumMusQA: A Human-written Music Understanding QA Benchmark Dataset},
  author={Weck, Benno and others},
  booktitle={Proceedings of the 4th Workshop on NLP for Music and Audio (NLP4MusA 2026)},
  pages={58--67},
  year={2026}
}

@inproceedings{humphrey2018openmic,
  title={Openmic-2018: An open data-set for multiple instrument recognition.},
  author={Humphrey, Eric and Durand, Simon and McFee, Brian},
  booktitle={Proceedings of the 19th International Society for Music Information Retrieval Conference (ISMIR)},
  pages={438--444},
  year={2018}
}

@ARTICLE{tzanetakis2002gtzan,
  author={Tzanetakis, G. and Cook, P.},
  journal={IEEE Transactions on Speech and Audio Processing}, 
  title={Musical genre classification of audio signals}, 
  year={2002},
  volume={10},
  number={5},
  pages={293-302},
  doi={10.1109/TSA.2002.800560}}

@article{turnbull2008semantic,
  title={Semantic annotation and retrieval of music and sound effects},
  author={Turnbull, Douglas and others},
  journal={IEEE Transactions on Audio, Speech, and Language Processing},
  volume={16},
  number={2},
  pages={467--476},
  year={2008},
  publisher={IEEE}
}

@article{li2026mrmad,
  title={Mrmad: A multi-round multi-audio benchmark for evaluating acoustic degradation perception in large audio-language models},
  author={Li, Yize and others},
  journal={arXiv preprint arXiv:2608.22236},
  year={2026}
}

@misc{pandora_mgp,
  author       = {{Pandora Media}},
  title        = {The Music Genome Project},
  howpublished = {\url{https://www.pandora.com/corporate/mgp.shtml}}
}

\end{document}